# Generation of bright laser-like coherent emissions from nitrogen molecular ions in intense infrared laser fields: a unique quantum system for extreme nonlinear optics

Jinping Yao[1], Wei Chu[1], Zhaoxiang Liu[1,2], Bo Xu[1,2], Jinming Chen[1,2,3], and Ya Cheng[1,4,5,*]

[1]*State Key Laboratory of High Field Laser Physics, Shanghai Institute of Optics and Fine Mechanics, Chinese Academy of Sciences, Shanghai 201800, China*

[2]*University of Chinese Academy of Sciences, Beijing 100049, China*

[3]*School of Physical Science and Technology, ShanghaiTech University, Shanghai 200031, China*

[4]*State Key Laboratory of Precision Spectroscopy, East China Normal University, Shanghai 200062, China*

[5]*Collaborative Innovation Center of Extreme Optics, Shanxi University, Taiyuan, Shanxi 030006, China*

[*]*ya.cheng@siom.ac.cn*

**Abstract**

We report on an investigation of simultaneous generation of several narrow-bandwidth laser-like coherent emissions from nitrogen molecular ions ($N_2^+$) produced in intense mid-infrared laser fields. With systematic examinations on the dependences of $N_2^+$ coherent emissions on gas pressure as well as several laser parameters including laser intensity, polarization and wavelength of the pump laser pulses, we reveal that the multiple coherent emission lines generated in $N_2^+$ originate from a series of nonlinear processes beginning with four-wave mixing, followed with stimulated Raman scattering. Our analyses further show that the observed nonlinear processes are greatly enhanced at the resonant wavelengths, giving rise to high conversion efficiencies from the infrared pump laser pulses to the coherent emission lines near the transition wavelengths between the different vibrational energy levels of ground $N_2^+(X^2\Sigma_g^+)$ and that of the excited $N_2^+(B^2\Sigma_u^+)$ states.

The advent of ultrashort and ultraintense laser pulses has revolutionized the interaction of light with matter, giving rise to non-perturbative tunnel ionization which is the pillar stone of strong field laser physics [1]. With the tunnel ionization as the initiative process, highly nonlinear processes such as high-order harmonic generation [2], above threshold ionization [3], and non-sequential double ionization [4] have been observed, which further provide the means to access the dynamics in atomic and molecular systems on attosecond time scale. Recently, lasing actions induced by tunnel ionization of nitrogen molecules have been observed which come as a major surprise to those who have been investigating strong field physics over the past three decades [5-21]. These observations were made with either a pump laser at 800 nm wavelength [6,8-20] or that at longer wavelengths in the range from 1 μm to 4 μm [5,7,21].

Interestingly, further pump-probe investigations on the generation of $N_2^+$ lasers at 391.4 nm, which corresponds to the transition between $N_2^+(B^2\Sigma_u^+, v'=0)$ and $N_2^+(X^2\Sigma_g^+, v=0)$ states, at the pump wavelengths clearly show that the gain of the external seed pulses whose spectra overlap the 391.4 nm transition line can only be observed for the pump laser at 800 nm wavelength but not the other wavelengths in the infrared range [18]. The experimental results lead to two consequences as follows. First, the nitrogen molecular ions produced in the 800 nm fields are preferentially populated on the excited $N_2^+(B^2\Sigma_u^+, v'=0)$ state but not the ground $N_2^+(X^2\Sigma_g^+, v=0)$ state, resulting in a population inversion which is difficult to understand in the framework of tunnel ionization [1,18]. Second, there should be no population inversion in nitrogen molecular ions produced in laser fields at infrared wavelengths ranging from 1 μm to 4 μm as evidenced by the failure of observing any gain in the external seed pulses injected immediately after the pump pulses to avoid temporal overlap between the two [18]. The origin of the laser-like emissions observed only with the self-generated seed pulses (i.e., the harmonics of the pump laser) at pump wavelengths longer than 1 μm is another puzzle yet to be unlocked.

To understand how the $N_2^+$ lasers are generated with the pump pulses of longer wavelengths, we perform a systematic investigation on the generation of $N_2^+$ lasers at various pump wavelengths in the range of 1.2 μm and 2 μm, which is the wavelength range of our OPA source. Surprisingly, our observations indicate that some conclusions drawn from the previous experiments are not always true. Among the deviations, the most significant two features reported here are (1) generation of the $N_2^+$ lasers can be achieved even when the spectra of seed pulses do *not* overlap the observed $N_2^+$ laser wavelengths, and (2) the laser lines can be significantly broadened to form a supercontinuum-like spectrum with a bandwidth of ~80 nm at low gas pressures. These findings provide vital clues for understanding the extreme nonlinear interactions between the strong laser fields and nitrogen molecular ions.

The experiments were carried out using an optical parametric amplifier (OPA, HE-TOPAS, Light Conversion Ltd.), which was pumped by a commercial Ti:sapphire laser system (Legend Elite-Duo, Coherent, Inc.). The OPA enables to generate wavelength-tunable femtosecond laser pulses in the range from 1.2 μm to 2.4 μm at a repetition rate of 1 kHz. The pump laser pulses were then focused into the gas chamber filled with nitrogen gas using an *f*=10 cm lens, and were further collimated using another lens with the same focal length. The signal beam exiting from the gas chamber together with the pump laser were attenuated, and the residual pump laser was eliminated with a piece of blue glass. At last, the generated third and fifth-order harmonic beams were focused into a grating spectrometer (Shamrock 303i, Andor) by a lens with a focal length of 15 cm. The spectra of the fundamental waves after passing through the gas chamber were measured by a fiber spectrometer (NIRQuest512, Ocean Optics, Inc.) with an integration sphere.

Figure 1(a) shows the dependence of the spectra of several $N_2^+$ emission lines on the gas pressure of nitrogen molecules obtained with pump pulses at a wavelength of 1580 nm. The pump power is approximately 1.0 W, and the beam diameter ($1/e^2$ width) is 7.6 mm. The pulse duration is measured to be ~60 fs. At the pump wavelength, neither the third nor the fifth harmonic of the pump laser could cover the transition lines at either 358.2 nm or 391.4 nm wavelengths. Surprisingly, two strong coherent emissions appear near the two transition lines at 358.2 nm and 391.4 nm wavelengths, which correspond to $N_2^+(B^2\Sigma_u^+, v'=1) \rightarrow N_2^+(X^2\Sigma_g^+, v=0)$ and $N_2^+(B^2\Sigma_u^+, v'=0) \rightarrow N_2^+(X^2\Sigma_g^+, v=0)$, respectively. Meanwhile, a weak emission at ~331 nm also appears on the fifth harmonic spectrum, which corresponds to the transition $N_2^+(B^2\Sigma_u^+, v'=2) \rightarrow N_2^+(X^2\Sigma_g^+, v=0)$. At high gas pressures above 40 mbar, the coherent emissions are mostly of narrow-bandwidths which is the typical feature in fifth harmonic generation in many gaseous media under our laser condition. The relatively high gas pressure will prevent the buildup of high peak intensity due to the plasma defocusing. However, at lower gas pressures in the range from 10 mbar to 30 mbar, pronounced supercontinuum spectra have been observed which fill up the gap between ~331 nm and ~358 nm as well as that between ~358 nm and ~391 nm. With the increase of the focal length of lens, the gas pressure for generating the supercontinuum decreases [22]. Naturally, one would expect that the supercontinuum generation near the transition lines of $N_2^+$ is a characteristic related to the energy level structure of molecular ions. We have confirmed this by carrying out a comparative experiment in argon, as argon has an ionization potential and a nonlinear coefficient similar to that of the nitrogen molecules [23]. In this case, all of the experimental conditions remained the same except that the gas in the chamber was changed from nitrogen molecules to argon atoms. The fifth harmonic spectrum generated in argon as a function of the gas pressure is shown in Fig. 1(b), which shows a smooth spectrum with a reasonable bandwidth of 9~10 nm (FWHM) and centered at the wavelength of ~319 nm (i.e., close to one fifth of the pump wavelength). The bright narrow-bandwidth emissions as well as the supercontinua both disappear in argon, as shown in Fig. 1(b). More quantitative comparison of fifth harmonic spectra in nitrogen molecules and argon atoms can be found in the Supplementary Material [22].

An immediate thought after seeing the spectrum in Fig. 1(a) would be that such unusual fifth harmonic spectra should be a result of the strong distortion in the spectrum of pump pulses and/or that of the third harmonic wave caused by nonlinear propagation of the pump laser. Therefore, we measured the spectra of the pump pulses and third harmonic in both nitrogen and argon gases at different gas pressures. As shown in Fig. 2(a-d), both the spectra in nitrogen and argon are very similar without any noticeable differences. From Fig. 2(a) and (c), one can see that below 30 mbar, the spectra of the pump pulses are undergoing a continuous blue shifting as well as a spectral broadening with the increase of the gas pressure. The blue shift can be attributed to the plasma generation as have been intensively discussed in previous investigations [24]. Above 30 mbar, the spectral broadening becomes more pronounced than the blue shift. For the third harmonic signals, the intensities increase with the increasing gas pressure, and the spectra broaden as well, as shown in Fig. 2(b) and (d). Nevertheless, both the pump laser at the fundamental wavelength and the generated third harmonic signal show limited increases in the spectral bandwidths which are much less pronounced than that of the fifth harmonic in Fig. 1(a). Thus, we exclude the possibilities of directly generating the supercontinuum via the fifth harmonic generation of the pump laser or its mixing frequency with the third harmonic.

To gain insight into the physical mechanism behind the result in Fig. 1(a), we further investigate the dependences of the fifth harmonic signals generated in nitrogen molecules on the intensity, polarization and wavelength of the pump laser. Figure 3(a) shows the supercontinuum spectra measured at different pump laser intensities for a fixed gas pressure of 16 mbar. The supercontinuum already began to appear at a relatively low pump power of 200 mW, corresponding to a peak intensity of $\sim 3 \times 10^{14}$ W/cm$^2$ as evaluated by assuming linear propagation of the pump laser in the gas chamber. When the pump laser intensity increased, the supercontinuum became stronger and its spectrum became broader. Then, we changed polarization of the pump laser at an average power of ~1 W from linear to circular polarization using an achromatic quarter-wave plate (QWP) at the gas pressure of 16 mbar. Here, the angle of 0º corresponds to linear polarization, and the angles of ±45º correspond to circular polarization. As shown by Fig. 3(b), the supercontinua generated with the linearly polarized pump lasers are the strongest in its intensity and the broadest in its spectrum. In contrast, changing the polarization of pump pulses from linear to circular polarization leads to a significant drop of the signal intensities by 3~4 orders of magnitude. Similar polarization dependence of the fifth harmonic, laser-like emissions and the supercontinuum shows that all these emission could share a similar physical mechanism.

Another important feature of the supercontinuum generation is its dependence on the pump wavelength. To clarify this, we examined the supercontinuum generation at two additional pump wavelengths of 1780 nm and 1400 nm, which are presented in Fig. 3(c) and (d), respectively. The pump powers at 1780 nm and 1400 nm were approximately 0.83 W and 1.0 W, respectively, which are the maximum output powers of our OPA system at the two wavelengths. The beam diameters were measured to be 9.3 mm and 6.2 mm for the 1780 nm and 1400 nm pump laser, respectively, and their pulse durations are close to that of 1580 nm (i.e., ~60 fs). It is noteworthy that the spectral intensities measured at different pump wavelengths are not calibrated. It was observed that at the pump wavelength of 1780 nm, the coherent emissions at ~391 nm and ~358 nm can be both efficiently generated as evidenced by Fig. 3(c), with a supercontinuum filling up the gap between the two emission lines. Although strong coherent emission at ~428 nm corresponding to the transition $N_2^+(B^2\Sigma_u^+, v'=0) \rightarrow N_2^+(X^2\Sigma_g^+, v=1)$ is also observed, no supercontinuum is observed between the laser lines at ~391 nm and ~428 nm. Meanwhile, some weak emissions from the transitions between the high vibrational energy levels of $N_2^+(B^2\Sigma_u^+)$ and that of $N_2^+(X^2\Sigma_g^+)$ states appears in the spectral range of 410~420 nm. In contrast, switching the pump wavelength from 1780 nm to 1400 nm leads to an almost complete elimination of the coherent emissions at ~391 nm and ~358 nm as well as the supercontinuum in between, as shown in Fig. 3(d). In such a case, only a strong emission at ~428 nm appears on the third harmonic spectrum. The results in Fig. 1(a), Fig. 3(c) and (d) indicate the important role of the pump wavelength in the generation of the supercontinuum and $N_2^+$ coherent emissions.

The key message from the results in Fig. 1 and Fig. 3 is that the $N_2^+$ coherent emissions at ~391 nm and ~358 nm wavelengths can only be achieved when the spectra of fifth harmonics overlap one of the transition lines between $N_2^+(B^2\Sigma_u^+)$ and $N_2^+(X^2\Sigma_g^+)$ states. This feature suggests that the fifth harmonic generation can be enhanced as a result of some resonant nonlinear processes. Although the fifth harmonic can be generated through either a direct fifth-order nonlinear process (i.e., $5\omega = \omega + \omega + \omega + \omega + \omega$) or a cascaded third-order process (i.e., $5\omega = 3\omega + \omega + \omega$), the latter should play a dominant role for our case [22]. The resonant four-wave mixing process is schematically depicted in Fig. 4(a). We would like to stress that due to the high peak intensity of pump pulses, the resonant process can be very efficient. As shown in Fig. 1(a), the resonantly generated fifth harmonic spectrum is about one order of magnitude stronger than that in argon, and the resonant process forces the central wavelength of the fifth harmonic to shift to near 331 nm.

Subsequently, the generated $5\omega$ photons at ~331 nm can initiate the stimulated Raman scattering (SRS) in the excited $N_2^+$ ions, which leads to the generation of photons at ~358 nm, as depicted in Fig. 4(b). Once the coherent emission at ~358 nm is generated, it triggers further SRS process to produce the photons at ~391 nm, as depicted in Fig. 4(c). In general, Raman scattering as illustrated in Fig. 4(b) and (c) requires an initial population in $N_2^+(B^2\Sigma_u^+)$ state [25]. In our experimental conditions (i.e., tight focusing geometry, high power and low pressure), the laser intensity reaches above $4\times10^{14}$ W/cm$^2$. According to ADK theory [26], more than 20% of neutral nitrogen molecules can be ionized directly from the inner orbital $\sigma_u 2s$ into excited $B^2\Sigma_u^+$ state under such a laser condition. Thus, the Raman processes in Fig. 4(b) and (c) is justified.

One concern is that the fifth harmonic generated in $N_2^+$ ions might be too weak to initiate SRS processes. To understand this, it should be noticed that the resonance of the Stokes photons generated by the Raman process in Fig. 4(b) and (c) with the electronic states of $N_2^+$ ions will efficiently promote the gain and meanwhile, reduce the threshold of SRS [27]. Moreover, the Raman process can be more efficient for the vibrational ground state because of the larger population, smaller damping rate, and larger dipole moment in the vibrational ground state. These effects together lead to the generation of stronger signal at ~391 nm. Some more rigorous analysis can be found in the Supplementary Material [22].

The last question is how the supercontinua between the stimulated Raman lasing lines are generated although they are not resonant with any transitions between $N_2^+(X^2\Sigma_g^+)$ and $N_2^+(B^2\Sigma_u^+)$ states. It is noteworthy that rotational transitions between the two electronic states cannot cover such broad range [9]. On the other hand, as we have discussed above, high peak intensities can be reached in our experimental condition. Therefore, cross phase modulation can occur immediately after the generation of the stimulated Raman lasers in the intense pump laser field, as illustrated in Fig. 4(d). In particular, the Raman lines near 358 nm and 391 nm wavelengths are in resonances with the transitions between $N_2^+(X^2\Sigma_g^+)$ and $N_2^+(B^2\Sigma_u^+)$ states. In such case, the nonlinear Kerr coefficient will be greatly enhanced, leading to the extraordinary supercontinuum generation which is rarely observed with atomic or neutral molecular gases under the similar experiment conditions [22]. In addition, because the cross phase modulation mainly occurs in the falling edge of the pump pulse, the spectral blue shift is much more pronounced than red shift [28]. The physics involved in the processes above is discussed more thoroughly in the Supplementary Material [22].

When the pump wavelength is switched to 1780 nm, a nonlinear process similar to that in Fig. 4(a-d) would dominate the generation for both the lines at ~358 nm and ~391 nm wavelengths as well as the supercontinuum in between. For the pump laser at 1400 nm, although its third harmonic spectrum can cover the transition line at ~428 nm, a low population in $X^2\Sigma_g^+(v=1)$ state makes it difficult to excite both SRS and cross phase modulation [22]. As a result, the supercontinuum emission is hardly observed at the pump wavelength, which is in good agreement with the experimental observation in Fig. 3(d).

To conclude, we have investigated near-resonant extreme nonlinear optics in nitrogen molecular ions at several wavelengths between 1.2 μm and 2 μm. Due to the existence of abundant vibrational energy levels, a series of third-order nonlinear optical processes including three-photon excitation, stimulated Raman scattering, and cross phase modulation can occur almost simultaneously at multiple resonant wavelengths. We notice that nitrogen molecular ions are a unique quantum system for demonstrating such extreme nonlinear effects. Since these molecular ions have a large ionization potential as high as ~28 eV, they can survive against photoionization at visible and infrared (IR) wavelengths even when the pump laser intensity reaches $8.5\times10^{14}$ W/cm$^2$ (assuming an ionization probability of ~1%). In the meantime, the nitrogen molecular ions can also allow the resonant transitions between various vibrational levels of $N_2^+(X^2\Sigma_g^+)$ and $N_2^+(B^2\Sigma_u^+)$ states to occur at near ultraviolet (UV) wavelengths, which can be resonant with the third or fifth harmonics of the IR pump laser. We envisage that the extreme nonlinear optics in molecular ions will not only become an intriguing topic of research for further exploration but also hold great promise for air-laser-based remote sensing applications.

This work is supported by the National Basic Research Program of China (Grant No. 2014CB921303), National Natural Science Foundation of China (Grant Nos. 61575211, 11674340, 61405220 and 11404357), Research Programs of the Chinese Academy of Sciences (Grant Nos. XDB16000000, QYZDJ-SSW-SLH010), and Shanghai Rising-Star Program (Grant No. 17QA1404600).

**References:**

[1] L.V. Keldysh, Zh. Eksp. Teor. Fiz. **47**, 1945 (1964) [Sov. Phys. JETP **20**, 1307 (1965)].

[2] A. L'Huillier and Ph. Balcou, Phys. Rev. Lett. **70**, 774–777 (1993).

[3] K. J. Schafer, Baorui Yang, L. F. DiMauro, and K. C. Kulander, Phys. Rev. Lett. **70**, 1599–1602 (1993).

[4] A. l'Huillier, L. A. Lompre, G. Mainfray, and C. Manus, Phys. Rev. A **27**, 2503–2512 (1983).

[5] J. Yao, B. Zeng, H. Xu, G. Li, W. Chu, J. Ni, H. Zhang, S. L. Chin, Y. Cheng, and Z. Xu, Phys. Rev. A **84**, 051802(R) (2011).

[6] Q. Luo, W. Liu, and S. L. Chin, Appl. Phys. B **76**, 337–340 (2003).

[7] D. Kartashov, J. Möhring, G. Andriukaitis, A. Pugžlys, A. Zheltikov, M. Motzkus, and A. Baltuška, CLEO:QELS-Fundamental Science, OSA Technical Digest QTh4E.6 (2012).

[8] J. Yao, G. Li, C. Jing, B. Zeng, W. Chu, J. Ni, H. Zhang, H. Xie, C. Zhang, H. Li, H. Xu, S. L. Chin, Y. Cheng, and Z. Xu, New J. Phys. **15**, 023046 (2013).

[9] H. Zhang, C. Jing, J. Yao, G. Li, B. Zeng, W. Chu, J. Ni, H. Xie, H. Xu, S. L. Chin, K. Yamanouchi, Y. Cheng, and Z. Xu, Phys. Rev. X **3**, 041009 (2013).

[10] Y. Liu, Y. Brelet, G. Point, A. Houard, and A. Mysyrowicz, Opt. Express **21**, 22791–22798 (2013).

[11] T. Wang, J. F. Daigle, J. Ju, S. Yuan, R. Li, and S. L. Chin, Phys. Rev. A **88**, 053429 (2013).

[12] J. Ni, W. Chu, C. Jing, H. Zhang, B. Zeng, J. Yao, G. Li, H. Xie, C. Zhang, H. Xu, S. L. Chin, Y. Cheng, and Z. Xu, Opt. Express **21**, 8746–8752 (2013).

[13] G. Point, Y. Liu, Y. Brelet, S. Mitryukovskiy, P. Ding, A. Houard, and A. Mysyrowicz, Opt. Lett. **39**, 1725–1728 (2014).

[14] B. Zeng, W. Chu, G. Li, J. Yao, H. Zhang, J. Ni, C. Jing, H. Xie, and Y. Cheng,

Phys. Rev. A **89**, 042508 (2014).

[15] Y. Liu, P. Ding, G. Lambert, A. Houard, V. Tikhonchuk, and A. Mysyrowicz, Phys. Rev. Lett. **115**, 133203 (2015).

[16] P. Wang, C. Wu, M. Lei, B. Dai, H. Yang, H. Jiang, and Q. Gong, Phys. Rev. A **92**, 063412 (2015).

[17] H. Xu, E. Lötstedt, A. Iwasaki, and K. Yamanouchi, Nature Commun. **6**, 8347 (2015).

[18] J. Yao, S. Jiang, W. Chu, B. Zeng, C. Wu, R. Lu, Z. Li, H. Xie, G. Li, C. Yu, Z. Wang, H. Jiang, Q. Gong, and Y. Cheng, Phys. Rev. Lett. **116**, 143007 (2016).

[19] L. Arissian, B. Kamer, and A. Rasoulof, Opt. Commun. **369**, 215–219 (2016).

[20] M. Lei, C. Wu, A. Zhang, Q. Gong, and H. Jiang, Opt. Express **25**, 4535–4541 (2017).

[21] A. Azarm, P. B. Corkum, and P. G. Polynkin, in Conference on Lasers and Electro-Optics, OSA Technical Digest (online) (Optical Society of America, 2016), paper JTh4B.9.

[22] See Supplemental Material at http://link.aps.org/supplemental/ for more details.

[23] V. Loriot, E. Hertz, O. Faucher, and B. Lavorel, Opt. Express **18**, 3011–3012 (2010).

[24] S. C. Rae and K. Burnett, Phys. Rev. A **46**, 1084–1090 (1992).

[25] D. C. Hanna, M. A. Yuratich, and D. Cotter, *Nonlinear Optics of Free Atoms and Molecules*, Springer-Verlag Berlin Heidelberg New York, 1979, pp. 187–194.

[26] M. V. Ammosov, N. B. Delone, and V. P. Krainov, Sov. Phys. - JETP **64**, 1191–1194 (1986).

[27] H. Yanagi, A. Yoshiki, S. Hotta, and S. Kobayashi, Appl. Phys. Lett. **83**, 1941–1943 (2003).

[28] P. L. Baldeck, R. R. Alfano, and G. P. Agrawal, Appl. Phys. Lett. **52**, 1939–1941 (1988).

**Captions of figures:**

Fig. 1 (Color online) Dependences of the fifth harmonic spectra generated in (a) nitrogen and (b) argon gases with the 1580 nm pump laser on gas pressure. Note that the color bars are in log scale.

Fig. 2 (Color online) Dependences of the spectra of (a) the pump pulses at 1580 nm wavelength after passing through nitrogen gas as well as (b) the generated third harmonic on the gas pressure. For comparison, the corresponding results obtained from argon gas with the same laser conditions are present in (c) and (d), respectively. The color bars are in log scale.

Fig. 3 (Color online) Dependence of the fifth harmonic spectra generated in nitrogen gas at a fixed gas pressure of 16 mbar on (a) intensity and (b) polarization of the 1580 nm pump laser, and the pressure-dependent spectra of the fifth harmonic generated at different wavelengths of (c) 1780 nm and (d) 1400 nm of the pump laser. The color bars are in log scale.

Fig. 4 (Color online) Schematic diagram of the physical mechanism. (a) The signal around 331 nm is generated with a resonant four-wave mixing. (b) The emission at ~358 nm is generated through stimulated resonance Raman scattering. (c) The Raman lasing line at ~358 nm triggers further stimulated resonance Raman scattering to produce the photons at ~391 nm. (d) The emission lines at ~358 nm and ~391 nm are spectrally broadened by a resonant cross phase modulation to produce the supercontinuum.

**Fig. 1**

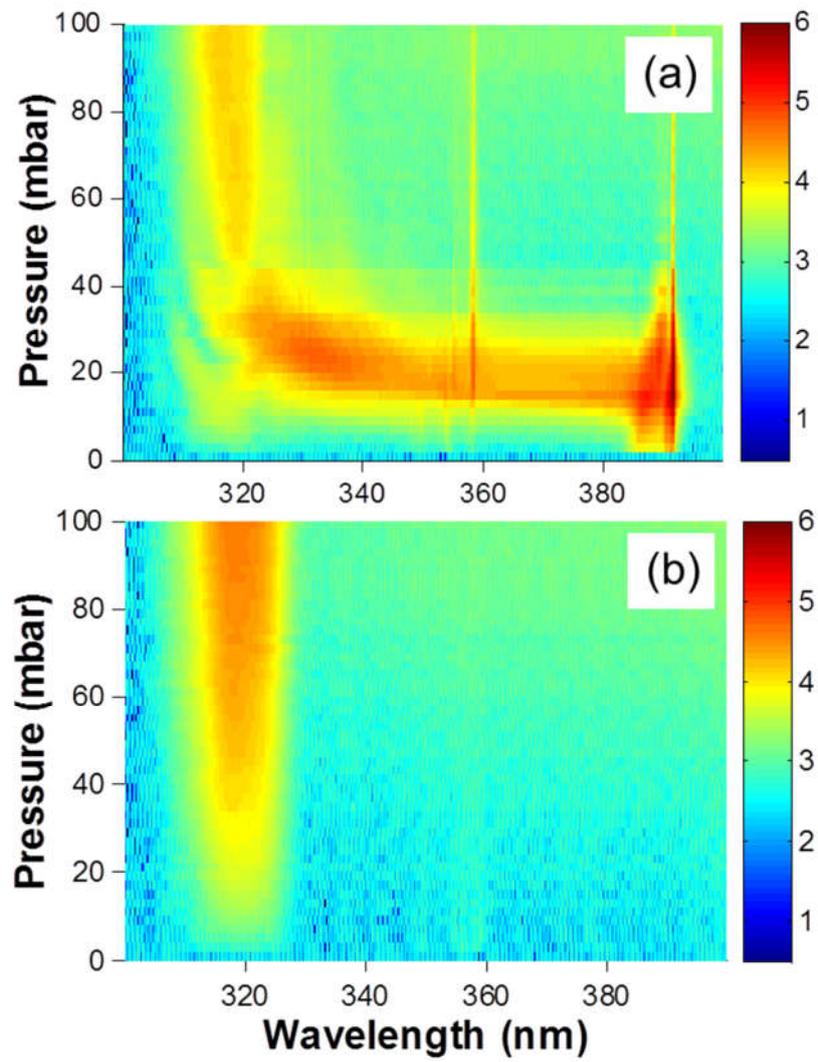

**Fig. 2**

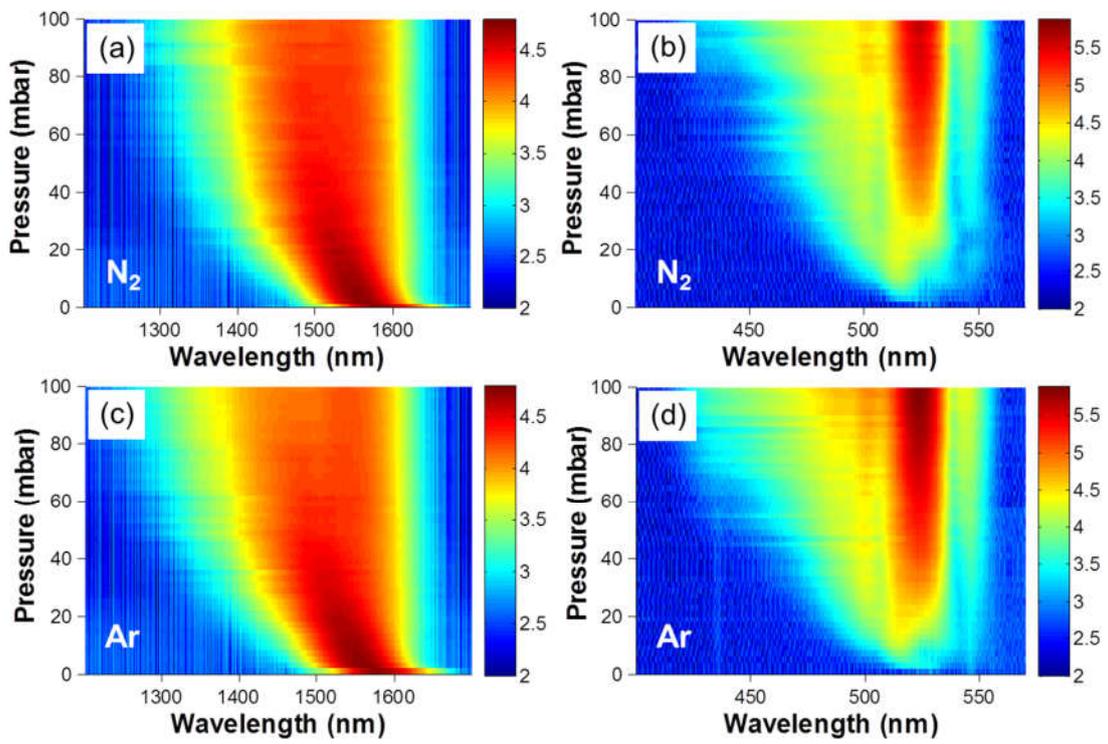

**Fig. 3**

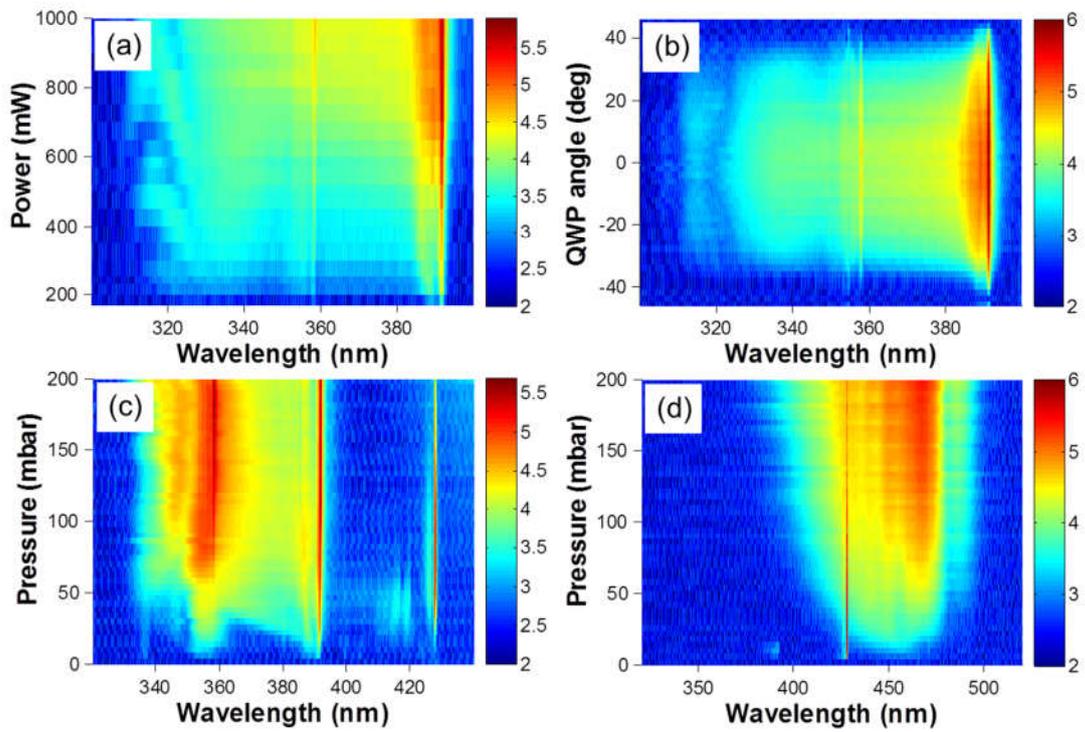

**Fig. 4**

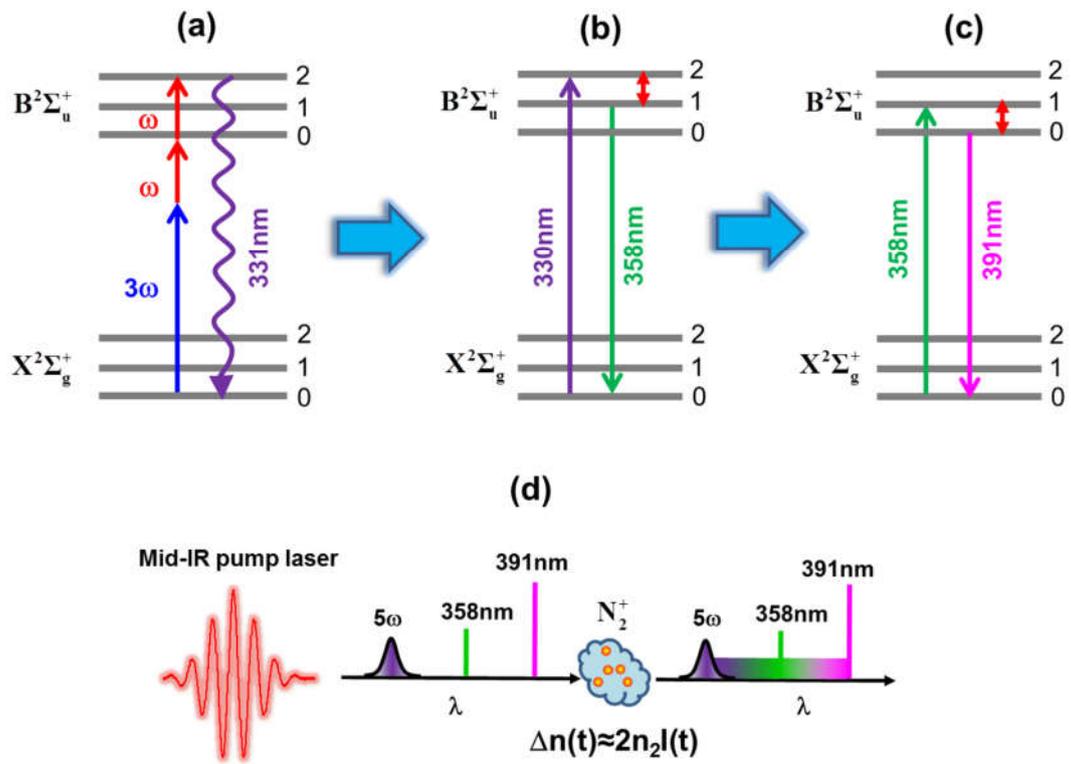

# Supplementary Material for

# Generation of bright laser-like coherent emissions from nitrogen molecular ions in intense infrared laser fields: A unique quantum system for extreme nonlinear optics

Jinping Yao[1], Wei Chu[1], Zhaoxiang Liu[1,2], Bo Xu[1,2], Jinming Chen[1,2,3], and Ya Cheng[1,4,5,*]

[1]*State Key Laboratory of High Field Laser Physics, Shanghai Institute of Optics and Fine Mechanics, Chinese Academy of Sciences, Shanghai 201800, China*
[2]*University of Chinese Academy of Sciences, Beijing 100049, China*
[3]*School of Physical Science and Technology, ShanghaiTech University, Shanghai 200031, China*
[4]*State Key Laboratory of Precision Spectroscopy, East China Normal University, Shanghai 200062, China*
[5]*Collaborative Innovation Center of Extreme Optics, Shanxi University, Taiyuan, Shanxi 030006, China*

[*]*ya.cheng@siom.ac.cn*

## 1. Quantitative comparison of fifth harmonic spectra in nitrogen and argon

Figure S1 compares the fifth harmonic spectra measured in nitrogen and argon at a gas pressure optimized for the observed strong supercontinuum generation (i.e., 20 mbar). The fifth harmonic signal in argon shows typical characteristics of harmonics generated in many gaseous media driven by strong laser fields, including a smooth spectrum with a bandwidth and central wavelength mainly determined by the pump pulses [S1]. In comparison with argon, the harmonic signal generated in nitrogen shows two striking differences. First, the fifth harmonic in nitrogen covers an extremely broad spectral range from ~310 nm to ~392 nm with two strong $N_2^+$ coherent emission lines at ~391 nm and ~358 nm superimposed on the supercontinuum spectrum. In particular, the coherent emission at ~391 nm is much stronger than other spectral components. Second, the fifth harmonic produced in nitrogen shows pronounced red-shift in its spectrum and enhancement of the signal intensity by approximately one order of magnitude in comparison with the fifth harmonic generated in argon. These differences are due to the different origins of harmonics, namely, the fifth harmonic generated in argon is mainly from the neutral atoms whereas the fifth harmonic generated in nitrogen molecules is mainly from the molecular ions.

It is suprising that the $N_2^+$ ion, which has an ionization potential of ~28 eV (in contrast, argon has an ionization potential only ~16 eV) [S2], can give rise to such a high conversion efficiency.

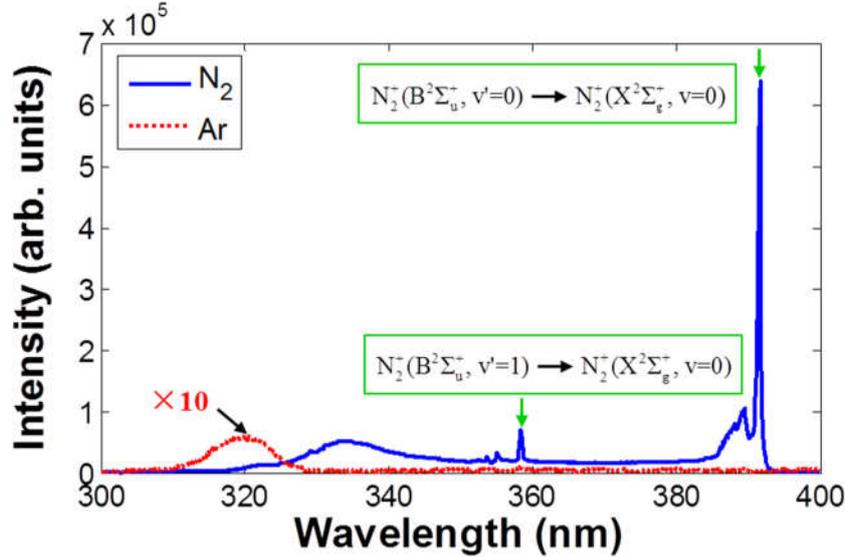

Fig. S1 Fifth harmonic spectra obtained in nitrogen and argon at the gas pressure of 20 mbar.

**2. Pressure-dependent supercontinuum emission at different focal geometries**

In the main text of our paper, Fig. 1(a) clearly shows that the supercontinuum generation is sensitive to gas pressure. To gain a better understanding, we also investigate the dependence of the supercontinuum emission on gas pressure with different focal conditions. When an $f$=10 cm lens is used, the supercontinuum reaches its maximum at ~20 mbar and can be observed in the relatively broad pressure range from 10 mbar to 30 mbar, as shown in Fig. 1(a). The optimum gas pressure for generating the supercontinuum decreases with the increase of the focal length. As presented in Fig. S2(a), the siginificant supercontinnum can be observed in the range of 6~10 mbar pressure with the $f$=15 cm focal lens. Lower gas pressures are required for the supercontinuum generation when the focal length of lens is further increased to 20 cm, as shown in Fig. S2(b). The result clearly indicates that high laser intensity is necessary for generating the supercontinuum. Thus, when the looser focal geometry is used, the gas pressure has to be reduced to maintain a sufficient laser intensity. The physics behind this observation will be discussed later.

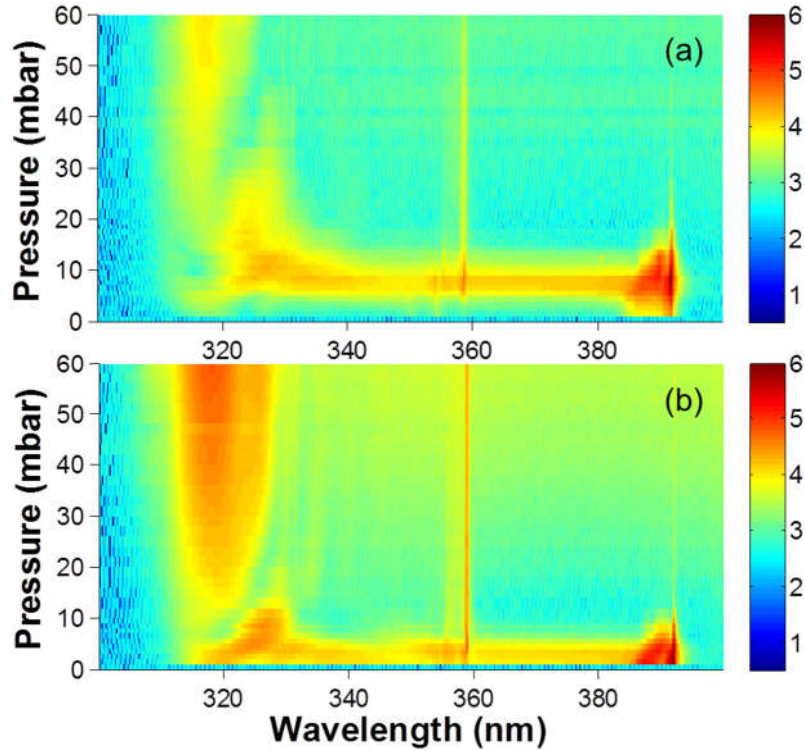

Fig. S2 Pressure-dependent fifth harmonic spectra in nitrogen with the focal lens of (a) $f$=15 cm and (b) $f$=20 cm.

### 3. Qualitative considerations on the physical picture

Based on the experimental results, we provide some qualitative considerations on the proposed physical mechanism which is responsible for generating the laser-like emission and the supercontinuum as well. As conceptually illustrated in Fig. 4 in the main text of our paper, the extraordinary spectral structure formed in nitrogen originates from a series of nonlinear processes beginning with four-wave mixing first in neutral nitrogen molecules and then in $N_2^+$ ions, followed by stimulated Raman scattering and cross phase modulation in $N_2^+$ ions. All these processes are greatly enhanced at the resonance wavelengths. We now discuss each process below.

#### 3.1 Resonant four-wave mixing

First, the comparison of the fifth harmonic spectra generated in nitrogen and argon shows that both the laser-like emission lines and broad-bandwidth supercontinuum depend on pump wavelengths, as their generation processes are related to the energy levels of $N_2^+(B^2\Sigma_u^+)$. In principle, the fifth harmonic spectrum of the pump laser at 1580 nm *cannot* cover the spectral lines at ~358 nm and ~391 nm, but it can cover the spectral line at ~331 nm, which corresponds to the transition $N_2^+(B^2\Sigma_u^+,v'=2) \rightarrow N_2^+(X^2\Sigma_g^+,v=0)$. In this case, the fifth harmonic generation will be enhanced due to multiphoton resonance. Although the fifth harmonic can be generated either through a direct fifth-order nonlinear process ($5\omega = \omega+\omega+\omega+\omega+\omega$) or a cascaded third-order process ($5\omega = 3\omega+\omega+\omega$), the latter should play a dominant role, as shown by N. A. Panov et al. [S3]. By examining the spectra of the fundamental wave and the third harmonic in Fig. 2, we can clearly see that the direct fifth-order nonlinear process can hardly generate the resonant $5\omega$ photons at the wavelength of ~331 nm due to the blue shift of the fundamental wave, whereas the cascade process can generate photons at the resonant wavelength thanks to the red-shifted third harmonic. On the other hand, the higher conversion efficiency can be achieved for third harmonic generation in neutral nitrogen molecules than $N_2^+$ ions for the non-resonant case. Thus, the conversion efficiency of fifth harmonic generation in the $N_2^+$ ions with the cascaded process will be strongly promoted thanks to the strong third harmonic generated up-stream in neutral molecules before the photoionization occurs near the focus.

The resonant four-wave mixing mentioned above is the initiative step in generating the laser-like emission lines and supercontinuum, which has been schematically depicted in Fig. 4(a). In the presence of resonance, its nonlinear susceptibility can be simplified as [S4]

$$\chi^{(3)}(-5\omega;3\omega,\omega,\omega) = \frac{N}{3\varepsilon_0 \hbar^3} \sum_{mn} \frac{\mu_{XB}\mu_{Bn}\mu_{nm}\mu_{mX}}{\omega_{BX}-5\omega-i\gamma_{BX}} \left[ \frac{1}{(\omega_{nX}-2\omega)(\omega_{mX}-\omega)} + \frac{1}{(\omega_{nX}-4\omega)(\omega_{mX}-\omega)} + \frac{1}{(\omega_{nX}-4\omega)(\omega_{mX}-3\omega)} \right] \quad (1).$$

Here, $\mu$ is transition matrix element and $\gamma$ dephasing rate. The subscripts X and B denote $N_2^+(X^2\Sigma_g^+)$ and $N_2^+(B^2\Sigma_u^+)$ states, and *m, n* are all possible intermediate states. The simplified expression ignores the contribution from nonresonant terms. From the formula, we can clearly see that at the resonant wavelength of ~331 nm (i.e., $\omega_{BX} - 5\omega \simeq 0$), the fifth harmonic generation is greatly enhanced. As shown in Fig. S1, the near-resonant four-wave mixing process forces the central wavelength of the fifth harmonic to shift to near 331 nm. The resonantly generated fifth harmonic spectrum is about one order of magnitude stronger than that in argon. It should be emphasized that it is a resonant nonlinear process in molecular *ions*, so that the pump laser must be sufficiently strong to generate a large amount of ions because otherwise the nonlinear optical signals generated in neutral molecules will smear the signals generated in the molecular ions. This is evidenced by the results obtained at the gas pressures above 40 mbar. The high gas pressure will produce too much plasma even at relatively low powers of pump pulse, which will prevent the buildup of high intensity due to the plasma defocusing. As shown in Fig. 1(a), the fifth harmonic spectra generated in nitrogen are similar to that in argon at relatively high gas pressures (i.e., > 40 mbar), implying that the fifth harmonic signal is mainly from neutral nitrogen molecules. As a result, we can separate the fifth harmonic signals from neutral molecules and that from molecular ions by tuning the gas pressure.

**3.2 Stimulated resonance Raman scattering**

We note that multiple spectral lines at ~331 nm, ~358 nm and ~391 nm correspond to the transitions from different vibrational energy levels of $N_2^+(B^2\Sigma_u^+)$ state to $N_2^+(X^2\Sigma_g^+, v=0)$ state. Subsequently, coherent emissions at ~358 nm and ~391 nm can be produced by a vibrational Raman scattering of molecular ions, as depicted in Fig. 4(b) and (c). This process is described by the expression below [S4,S5]

$$\chi_R^{(3)}(-\omega_s;\omega_p,-\omega_p,\omega_s)$$
$$= \frac{N/6\varepsilon_0\hbar^3}{(\Omega_{fg}+\omega_s-\omega_p+i\gamma_{fg})}\sum_i(\frac{\mu_{gi}\mu_{if}}{\Omega_{ig}-\omega_p+i\gamma_{ig}}+\frac{\mu_{gi}\mu_{if}}{\Omega_{ig}+\omega_s+i\gamma_{ig}})\sum_i(\frac{\mu_{fi}\mu_{ig}}{\Omega_{ig}-\omega_p-i\gamma_{ig}}+\frac{\mu_{fi}\mu_{ig}}{\Omega_{ig}+\omega_s+i\gamma_{ig}}) \quad (3).$$

Here, μ is transition matrix element, γ dephasing rate, $\Omega_{fg}$ energy difference of *f* and *g* states, $\omega_s$ frequency of the Stokes wave, and $\omega_p$ frequency of the pump wave. The *g*, *f* and *i* denote the initial, final energy levels and all possible intermediate states of Raman scattering, respectively.

From the formula, we note that Raman scattering can be enhanced for either $\omega_p \sim \Omega_{ig}$ or $\omega_s \sim -\Omega_{ig}$. The former will induce an oscillating optical polarizability of $N_2^+$ ions in $X^2\Sigma_g^+$ state with the frequency of $\Delta v=1$ and generate Stokes photons at the wavelength of ~356 nm, as shown in Fig. S3(a). The latter will excite an oscillation of $N_2^+$ ions in $B^2\Sigma_u^+$ state with the frequency of $\Delta v'=1$ and generate Stokes photons at the wavelength of ~358 nm, as presented in Fig. S3(b). Although the signal at ~356 nm wavelength corresponding to the transition $N_2^+(B^2\Sigma_u^+, v'=2) \rightarrow N_2^+(X^2\Sigma_g^+, v=1)$ is also observed in Fig. S1, it is much weaker than that at ~358 nm. Thus, Raman scattering from molecular ions in the excited state plays a dominant role here.

In general, the Raman scattering as illustrated in Fig. S3(b) requires an initial population in $N_2^+(B^2\Sigma_u^+, v'=1)$ state [S5]. We notice that our experiments were carried out under the tight focusing condition. Together with the high pulse energies and low gas pressures, the laser intensity can reach above $4\times10^{14}$ W/cm$^2$. Thus, according to ADK theory [S6], more than 20% molecules can be tunnel ionized from the inner $\sigma_u 2s$ orbital to generate ions in $B^2\Sigma_u^+$ state, and these ions will mainly populate in low vibrational energy levels of v'=0,1 [S7]. In addition, for the two resonant processes in Fig. S3, the difference in the term $\mu_{fi}\mu_{ig}$ is negligible since Franck-Condon factors for the two transitions $B^2\Sigma_u^+(v'=2) \rightarrow X^2\Sigma_g^+(v=1)$ and $B^2\Sigma_u^+(v'=1) \rightarrow X^2\Sigma_g^+(v=0)$ are close to each other [S8]. Based on this consideration, the stronger emission observed at ~358 nm than that at ~356 nm could be due to its smaller damping rate. The damping rate $\gamma_{ig}$ usually follows $\gamma_{ig} \sim 1/2 \cdot (1/\tau_i + 1/\tau_g)$, where $\tau_i$, $\tau_g$ indicate the lifetime of the level $i$ and the level $g$, respectively [S4]. For the process in Fig. S3(a), the $\gamma_{ig}$ is determined by the lifetimes of $X^2\Sigma_g^+(v=0)$ and $B^2\Sigma_u^+(v'=2)$ states, whereas the $\gamma_{ig}$ depends on the lifetimes of $X^2\Sigma_g^+(v=0)$ and $B^2\Sigma_u^+(v'=1)$ states for the process in Fig. S3(b). Thus, their difference in the damping rate is mainly from different vibrational energy levels of $B^2\Sigma_u^+$ state. Generally speaking, the lower vibration energy level (e.g., the level $B^2\Sigma_u^+(v'=1)$) is more stable than the higher one (e.g., the level $B^2\Sigma_u^+(v'=2)$) and thus has the smaller damping rate [S9]. This could be the main reason for the resonant process in Fig. S3(b) prevails against that in Fig. S3(a).

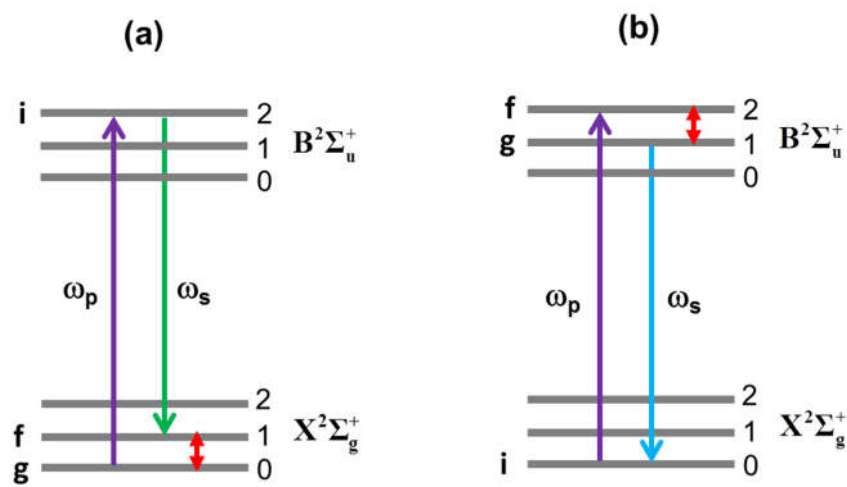

Fig. S3 Two types of resonance enhanced Raman scattering. (a) The pump photon is resonant with the intermediate *i* state lying above the initial *g* state. (b) The Stokes photon is resonant with the intermeidiate *i* state lying below the initial *g* state.

Intuitively, stimulated Raman scattering (SRS) usually requires a high intensity of the pump laser, whereas the fifth harmonic generated in the molecular ions might be too weak to initiate SRS processes. To understand this, it should be noticed that the resonance of the new photons generated by the Raman process in Fig. S3(b) with the transitions between the electronic states of $N_2^+$ ions will efficiently promote the gain and reduce the threshold of SRS [S10]. Thus, in the scheme of traveling wave amplification, the energy of fifth harmonic can be efficiently transferred to the photon at ~358 nm via stimulated resonance Raman scattering. The subsequent SRS with the ~358 nm as the pump will produce the photon at ~391 nm. This process is more efficient because of higher population [S7], smaller damping rate in the vibrational ground state

[S9], as well as larger dipole moment for the transition $B^2\Sigma_u^+(v'=0) \to X^2\Sigma_g^+(v=0)$ [S8]. In the meantime, the molecular oscillation with a frequency of $\Delta v'=1$ will be coherently excited by SRS from $B^2\Sigma_u^+(v'=2)$ to $B^2\Sigma_u^+(v'=1)$, which further reinforce the Raman process from $B^2\Sigma_u^+(v'=1)$ to $B^2\Sigma_u^+(v'=0)$. This mechanism is similar to that of the coherent Stokes Raman scattering [S4]. The effectiveness of this model is evidenced by the fact that the spectral line at ~391 nm is observed to be much stronger than the spectral line at ~358 nm.

**3.3 Resonant cross phase modulation**

The last question is how the supercontinua between the stimulated Raman lasing lines are generated although the supercontinua are not resonant with any transitions between the ground $N_2^+(X^2\Sigma_g^+)$ and excited $N_2^+(B^2\Sigma_u^+)$ states. It should be pointed out that the rotational transitions between two electronic states cannot cover such broad range [S11]. Spectral broadening is usually caused by the time-dependent refractive index in the nonlinear propagation. When the weak probe light (e.g., low-order harmonics, laser-like emission, etc.) propagates in the nitrogen gas together with the strong pump, cross-phase modulation induced by the pump will result in spectral broadening of the probe, which can be calculated by [S12]

$$\Delta\omega \approx \frac{2\omega_0 n_2 I}{c\tau} L \qquad (3).$$

Here, the nonlinear Kerr coefficient $n_2$ is proportional to the real part of the third order susceptibility. Because the weak probe light with the frequency of $\omega_0$ (i.e., the laser-like emission line at either ~358 nm or ~391 nm wavelengths) is resonant with the transitions between the electronic states of $N_2^+$, $n_2$ can be simplified as [S4]

$$n_2 \propto \text{Re}\{\chi^{(3)}(-\omega_0;\omega_0,\omega_{pu},-\omega_{pu})\}$$
$$= \text{Re}\left\{\frac{N}{6\varepsilon_0\hbar^3(\omega_{BX}-\omega_0-i\gamma_{BX})^2}\sum_n \mu_{XB}\mu_{Bn}\mu_{nB}\mu_{BX}\left(\frac{1}{\omega_{nX}-\omega_0-\omega_{pu}}+\frac{1}{\omega_{nX}-\omega_0+\omega_{pu}}\right)\right\} \quad (4).$$

Here, $\mu$ is transition matrix element, $\gamma$ dephasing rate, $\omega_{pu}$ frequency of the pump laser at the mid-infrared wavelength, $\omega_0$ frequency of the probe light. The $n$ denotes all possible intermediate states. From the formula, we can clearly see that $n_2$ will be enhanced greatly when $\omega_{BX} \approx \omega_0$. The condition is readily fulfilled for the coherent emission lines at ~358 nm and ~391 nm. Next, let us estimate whether the cross phase modulation can enable to induce an extremely broad spectral broadening required by the supercontinuum generation. Assuming that the pump intensity $I \approx 4\times10^{14}$ W/cm$^2$, the pulse duration $\tau \approx 60$ fs, the interaction length $L \approx 1$ cm, the supercontinuum generation from ~391 nm to ~358 nm requires $n_2 \geq 2.09\times10^{19}$ cm$^2$/W. The value is about two orders of magnitude higher than that of neutral nitrogen molecules at a gas pressure of 20 mbar [S13]. With this, we deduce that at the resonance wavelengths, the nonlinear Kerr coefficient of $N_2^+$ ions could be much higher than that of the neutral nitrogen molecules. It should be mentioned that since currently there is no report on the nonlinear optical coefficient in $N_2^+$ ions, our result provides the first experimental evidence on the strong nonlinear property of such quantum systems. In addition, because the cross phase modulation in $N_2^+$ ions mainly occurs in the falling edge of the pump pulse (i.e., most ions are produced after the crest of the pump pulses), the spectral blue shift is much more pronounced than red shift [S14]. The process is schematically depicted in Fig. 4(d).

Based on the analyses above, we can clearly see that the laser-like emission and extraordinary supercontinuum generation should be attributed to multiple resonant nonlinear processes in $N_2^+$ ions. The resonant interaction of molecular ions with the intense pump laser allows for decent conversion efficiencies for these nonlinear processes. When the pump wavelength is switched from 1580 nm to 1780 nm, a nonlinear process similar to that in Fig. 4(a-d) would dominate the generation of both the lines at ~358 nm and ~391 nm wavelengths as well as the supercontinuum in between. For the pump laser at 1400 nm, although its third harmonic spectrum can cover the transition line at ~428 nm, a low population in $X^2\Sigma_g^+(v=1)$ state [S7] makes it difficult to generate anti-Stokes photon at ~391 nm. Meanwhile, the low population will also lead to the decrease of nonlinear coefficient $n_2$. As a result, the supercontinuum emission is hardly observed with the pump wavelength of 1400 nm, which is in good agreement with the experimental result in Fig. 3(d).

**References:**

[S1]    A. V. Mitrofanov, A. A. Voronin, D. A. Sidorov-Biryukov, A. Pugžlys, E. A. Stepanov, G. Andriukaitis, T. Flöry, S. Ališauskas, A. B. Fedotov, A. Baltuška, and A. M. Zheltikov, Sci. Rep. **5**, 8368 (2015).

[S2]    F. H. Dorman and J. D. Morrison, J. Chem. Phys. **35**, 575 (1961).

[S3]    N. A. Panov, D. E. Shipilo, V. A. Andreeva, O. G. Kosareva, A. M. Saletsky, H. Xu, and P. Polynkin, Phys. Rev. A **94**, 041801(R) (2016).

[S4]    R. W. Boyd, *Nonlinear Optics*, third edition (Elsevier Pte Ltd., Singapore, 2009).

[S5]    D. C. Hanna, M. A. Yuratich, and D. Cotter, *Nonlinear Optics of Free Atoms and Molecules*, Springer-Verlag Berlin Heidelberg New York, 1979, pp. 187–194.

[S6]    M. V. Ammosov, N. B. Delone, and V. P. Krainov, Sov. Phys. - JETP **64**, 1191 (1986).

[S7]    A. Becker, A. D. Bandrauk, and S. L. Chin, Chem. Phys. Lett. **343**, 345 (2001).

[S8]    A. Lofthus and P. H. Krupenie, J. Phys. Chem. Ref. Data **6**, 113 (1977).

[S9]    J. Jolly and A. Plain, Chem. Phys. Lett. **100**, 425 (1983).

[S10]   H. Yanagi, A. Yoshiki, S. Hotta, and S. Kobayashi, Appl. Phys. Lett. **83**, 1941 (2003).

[S11]   H. Zhang, C. Jing, J. Yao, G. Li, B. Zeng, W. Chu, J. Ni, H. Xie, H. Xu, S. L. Chin, K. Yamanouchi, Y. Cheng, and Z. Xu, Phys. Rev. X **3**, 041009 (2013).

[S12]   R. R. Alfano, *The Supercontinuum Laser Source*, second edition (Springer Science+Business Media, Inc., 2006), pp. 122.

[S13]   V. Loriot, E. Hertz, O. Faucher, and B. Lavorel, Opt. Express **18**, 3011 (2010).

[S14]   P. L. Baldeck, R. R. Alfano, and G. P. Agrawal, Appl. Phys. Lett. **52**, 1939 (1988).